\documentclass[baaa]{baaa}

\usepackage[pdftex]{hyperref}
\usepackage{subfigure}
\usepackage{natbib}
\usepackage{helvet,soul}
\usepackage[font=small]{caption}

\contriblanguage{1}

\contribtype{4}

\thematicarea{11}

\title{On the origin of stellar associations}
\subtitle{The impact of \textit{Gaia} DR2}

\titlerunning{OB associations}

\author{G. Carraro\inst{1}}
\authorrunning{Carraro, G.}

\contact{giovanni.carraro@unipd.it}

\institute{
Dipartimento di Fisica e Astronomia Galileo Galilei, Padova, Italy
}

\resumen{}

\abstract{
In this review I discuss different theories of the formation of OB associations
in the Milky Way, and provide the observational evidences in support of them. 
In fact, the second release of \textit{Gaia} astrometric data (April 2018) is
revolutionising the field, because it allows us to unravel  the 3D structure and
kinematics of stellar associations with  unprecedented details by providing
precise distances and a solid membership assessment.  
As an illustration, I summarise some recent studies on three OB associations:
Cygnus OB2, Vela OB2, and  Scorpius OB1, focussing in more detail to Sco OB1.  A
multi-wavelength study, in tandem with astrometric and kinematic data from
\textit{Gaia} DR2, seems to lend support, at least in this case, to a scenario
in which star formation is not monolithic. As a matter of fact, besides one
conspicuous star cluster, NGC 6231, and the very sparse star cluster Trumpler
24, there are several smaller groups of young OB and pre-main sequence stars across the association, indicating that star formation is highly structured and with no preferred scale.  A new revolution is expected with the incoming much awaited third release of \textit{Gaia} data.
}


\begin{document}

\maketitle

\section{Introduction}
\label{S_intro}

Stellar associations \citep[a term introduced by][]{A1947} are unbound loose ensembles of young stars \citep[see][for an exhaustive recent review on this subject]{goul2018}. They are dominated by OB stars, which are so bright to be often visible
even at optical wave-lengths, in spite of the patchy and obscured environment in
which they form and spend most of their short life. As expected, they are rich
in pre-main sequence stars \citep{Damiani} as well. Being very young, they
should keep a vivid memory of the star formation (SF) event from which they
formed and therefore are ideal tracers to study different modes of SF. 
It is unclear whether stellar (OB or T) associations are a direct product of star formation, or, more conservatively, an intermediate dynamical stage during the early evolution of young star clusters, soon after their formation, and before their final dissolution.

In the first scenario the actual spatial distribution of stars is the one at birth. Since they do not show any special shape, 
SF is expected to occur at any scale, from very large scale, as large as a galaxy, down to scales as small as individual young stellar objects. SF is then hierarchical, and the stellar structures which emerge from it are often fractally organized. \citet{efre1989} identifies associations, aggregates, complexes and super-complexes. There is no preferred scale, and whatever density peak in the interstellar medium (ISM) can generate stars, from cluster size all the way to even individual stars.

In the second scenario, stars form in star clusters \citep{Lada03} which are
initially embedded, and made of 50 stars or more. They are formed then in the
upper part of the molecular cloud mass distribution. Then stellar winds, UV
flux, and eventually SNe explosions from the most massive stars in the pristine
cluster remove the gas not processed by SF and, in this way, move the star
cluster out of virial equilibrium. The cluster becomes loose and turns into an
OB association. Associations in this scenario are then the final stage of the evolution of star
clusters before dissolution in the general Galactic field. Being the
time scale of this process short compared with the average age of Milky Way open
clusters, this is normally referred to it as cluster ``infant
mortality''.  From an observational point of view, the 3D kinematics of stars should
show some indication of global expansion, in other words, some relationships
between radial velocity and radius. 

As mentioned in a recent study by \citet{Ward2019}, this monolithic scenario does not seem to be supported by \textit{Gaia} DR2 \citep{Gaia} data.  These authors do not find any significant correlation
between radial velocity and radius in
most OB associations which, therefore, do not show dominant expansion signatures confirming earlier results by \citep{Mel2018}. Instead, they seem to be  dominated by other large scale motions, which indicate that the velocity field is highly structured and cannot be reconciled with a monolithic collapse. 
To cast additional light on this topic, ideally one should study in detail many OB associations, coupling multi-wavelength campaigns with kinematical and astrometric data from \textit{Gaia} DR2.
Studies of this type have been  recently performed for some associations, like
Cygnus  OB2 \citep{Berla}, Vela OB2 \citep{Beccari, Canta} and Scorpius OB1
\citep{Damiani,Yalialeva2020}, to give a few
examples.  In all cases, as I will outline below in more detail, the evidence is
that SF is highly structured. 
 
\section{Recent results from \textit{Gaia} DR2 data exploitation}

The second release of \textit{Gaia} astrometric data prompted an intense investigation of stellar associations in the Milky Way.  I provide here a couple of illuminating examples of recent studies: Cygnus OB2 and Vela OB2. A full section will then be devoted to Scorpius OB1, in the study of which I was personally involved.

\subsection{Cygnus OB2}

 \citet{Berla} recently studied the young OB association Cygnus OB2 \citep{Wright} and found that this association is composed of two spatially separated, coeval, groups of stars. One main group is located at about 1760~pc, while a second, less prominent, group is located closer, at about 1350~pc. Besides, they find that the bulk of the association is more distant than previously thought.  This points to a scenario in which SF occurred at the same time but with different intensity in at least two different, detached, regions of the association.
 
\subsection{Vela OB2}
 
More interesting is the case of  the Vela OB2 association presented in \citet{Beccari}. Following up previous suggestions \citep{Jeff,Damiani17} that Vela OB2 is composed of two distinct stellar populations, \citet{Beccari} identified six independent physical groups with different distances, ages and kinematics (see Fig.~1).   This identification was done using photometry with the wide field camera OmegaCAM mounted at the VLT survey telescope on Paranal (Chile), spectroscopy from the \textit{Gaia}-ESO survey \citep{GaiaESO}, and parallaxes and proper motion components from \textit{Gaia} DR2 \citep{Gaia}. 
The various groups can be assigned to two different populations.  Four of them  (they are named Cl~1, Cl~2, Cl~4 and Cl~5) are 10~Myr old, while  the other two, Cl~3 and Cl~6,  are older (30~Myr). Cl~3 coincides with the young star cluster NGC~2547, and Cl~4 with the very young star cluster $\gamma$ Velorum, two very well known and widely studied star clusters. The other 4 are newly discovered physical groups. NGC~2547 (Cl~3) would be  located at a distance of 393~pc,  while $\gamma$~Velorum (Cl~4) would be at a distance of 391~pc, therefore closer than NGC~2547. As a consequence, the two different populations are separated by about 40~pc. The other groups, although sharing similarities in age either with NGC~2547 or $\gamma$ Velorum, are located at different distances and therefore they are not associated with the two more prominent clusters. In conclusion, Vela OB2 is quite a complex OB association. It is very difficult to provide a comprehensive picture since radial velocities are available only for a small group of stars preventing a full 3D characterisation of the association.  The authors finally underline that by tracing back the velocity vector it is not possible to find a common spatial origin for the six groups.
Nonetheless, it seems that the origin of the Vela OB2 association is linked to
the so-called \textit{IRAS} Vela shell, which defines the edge of the association  and it
is made of dust  and gas and with which the various detected stellar groups seem
to be associated. 
According to \citet{Canta}, who looked in detail at the spatial distribution and kinematics of Vela OB2, a SN event occurred before the formation of Vela OBs triggered a SF burst which eventually generated it. The SN in this scenario was a massive star from the 30~Myr population whose main representative is NGC~2547. This is a reasonable and appealing scenario, which can be confirmed once radial velocities will be available for a much larger fraction of stars. These, together with more precise age estimates, would allow to integrate back the orbit of the stars and search for a possible common birthplace.
In the case that such a common birthplace is found, this would support the proposed SN expansion mechanism.

\begin{figure}[!t]
  \centering
  \includegraphics[width=0.95\columnwidth]{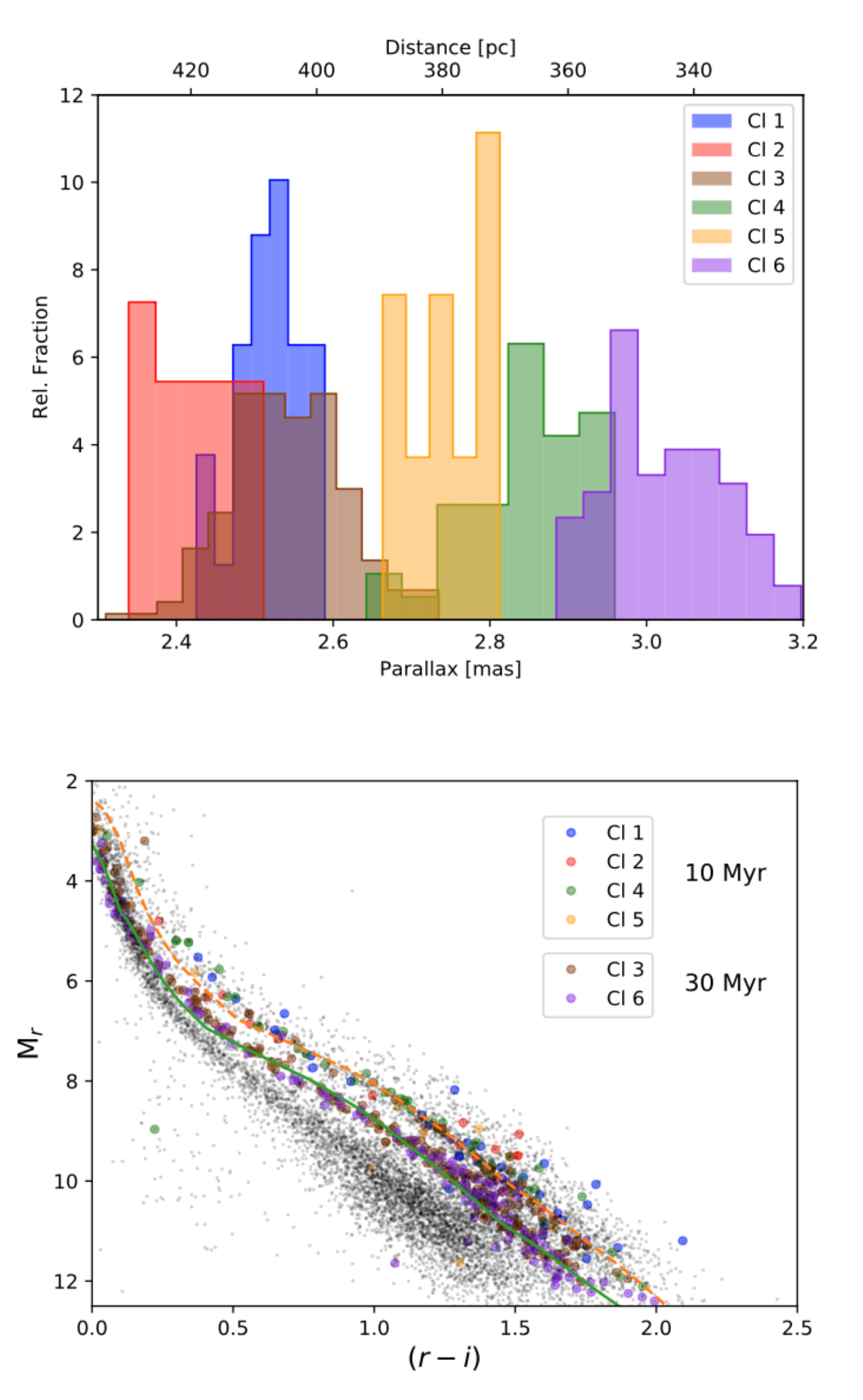}
  \caption{Stellar aggregates identified in the Vela OB2 associations by
\citet{Beccari}. They have different heliocentric distances and form two age
groups. Taken from \textit{A sextet of clusters in the Vela OB2 region revealed by Gaia}, \cite{Beccari}, Oxford University Press.} 
  \label{Vela}
\end{figure}

\def\q{\phantom{0}}
	\begin{table*}
		\caption{Fundamental parameters of the detected groups in the
Sco1 OB association by \citet{Yalialeva2020}. $R_V$ indicates the
ratio of total to selective absorption in the direction of each group.} 
		\label{tab:t4}
		\begin{center}
			\begin{tabular}{c c c c c c c c}
				\hline
\noalign{\smallskip}
				Group & $E(B-V)$  & $\sigma_{E(B-V)} $ &  Distance                &   log(Age)& $R_V $ & $\sigma_{R_{V}} $ &  $A_{V}$  \\ 
				& mag      & mag                 & pc                       &    dex    &        &                   &           \\ 
\noalign{\smallskip}
				\hline                                                                                                                    
\noalign{\smallskip}
				A     & 0.55      &  0.7               & $ 1608_{-35}^{+36\q} $   &   8.75    & 3.0    &  0.20              &   1.650    \\ 
				B1    & 0.36      &  0.1               & $ 1549_{-47}^{+181}  $   &   6.50   & 3.2    &  0.20              &   1.152   \\ 
				B2    & 0.39      &  0.3               & $ 1644_{-78}^{+82\q }$   &   6.75    & 3.1    &  0.20              &   1.209   \\ 
				B3    & 0.38      &  0.2               & $ 1629_{-21}^{+44\q }$   &   6.95    & 3.5    &  0.20              &   1.330    \\ 
				C     & 0.57      &  0.3               & $ 1578_{-11}^{+85\q }$   &   6.85    & 2.8    &  0.25             &   1.596   \\ 
				D     & 0.58      &  0.3               & $ 1761_{-36}^{+139}  $   &   8.00   & 3.1    &  0.25             &   1.798   \\ 
				E     & 0.57      &  0.3               & $ 1249_{-58}^{+61\q }$   &   8.15    & 3.2    &  0.25             &   1.824   \\ 
				F     & 0.47      &  0.1               & $ 1682_{-54}^{+137}  $   &  \llap{$>$}$6.60$   & 2.9    &  0.20              &   1.363   \\ 
				G     & 0.40     &  0.9               & $ 1524_{-54}^{+135}  $   &   6.70   & 2.5    &  0.20              &   1.000     \\ 
				
\noalign{\smallskip}
				\hline
			\end{tabular}
		\end{center}
	\end{table*}
	
\section{The OB association Sco 1}
Recent studies on this OB association have been performed by \citet{Damiani,
Kuhn}, and \citet{Yalialeva2020}, using \textit{Gaia} DR2 and
multi-wavelength photometry.  
Sco~1 is a very rich and complex stellar association \citep{Damiani}.
The spectacular  H\,\textsc{ii} region G345.45+1.50  is situated in the northern part of the field, while the most prominent young star cluster, NGC 6231, is located in the southern part. 
I will summarise here the results of \citet{Yalialeva2020}, who are not covering NGC~6231 (see Fig.~2), but  concentrate on the northern and central region.
Here, the most interesting structure  is  certainly Trumpler~24. This is thought
to be a young  open cluster with poorly defined boundaries and  complex
structure belonging to Sco~OB1 \citep{Heske}.  
Besides, the area under investigation is rich in pre-main sequence stars \citep{Heske, Damiani}, which indicates active/recent star formation.

The area shown in Fig.~2 was surveyed with multi-colour  \textit{UBVI} photometry \citep{Yalialeva2020} from the Las Campanas Henrietta Swope 1.0-meter telescope. Then, by cross-correlating with \textit{Gaia} DR2 data, the existence of different physical stellar groupings was searched for.

The clustering algorithm adopted is based on  the DBSCAN (Density-Based Spatial
Clustering of Applications with Noise) technique. The algorithm uses the
clustering module of the machine learning library \textsc{scikit-learn}
\citep{scikit-learn} as its basis.  Nine different groups were found and
characterised (see Table~1 and Fig.~3) in some detail, depending on the number
of recovered members.

\begin{figure}[!t]
  \centering
  \includegraphics[width=\columnwidth]{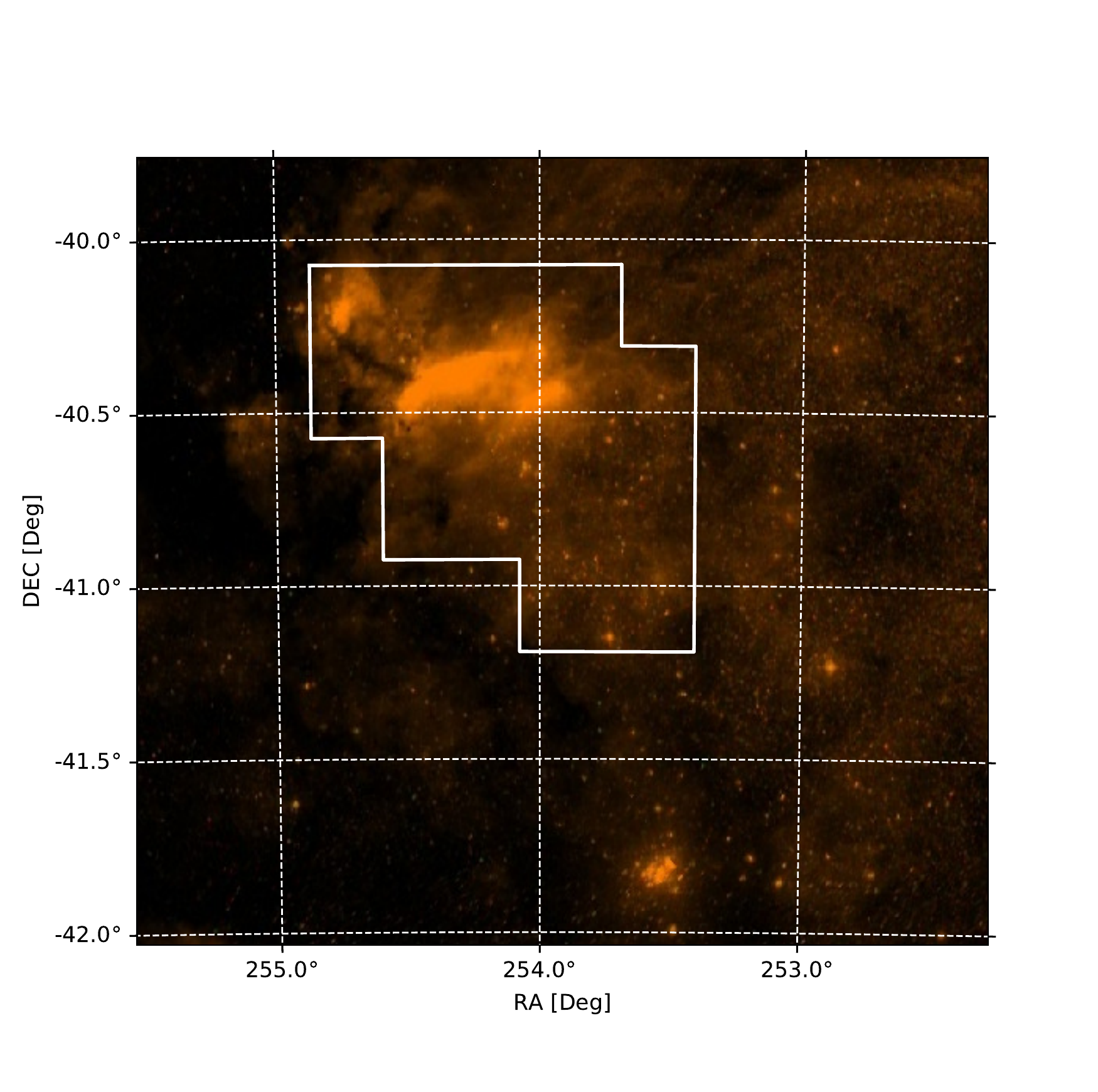}
  \caption{A DSS map of the region of the Sco~1 association. The white area is the one covered by \textit{UBVI} photometry. It includes Trumpler~24 in its totality,
the bright H\,\textsc{ii} region mentioned in the text and several other groups. The prominent cluster in the south-west corner is NGC 6231. Taken from \textit{A new look at Sco OB1 association with Gaia DR2}, \cite{Yalialeva2020}, Oxford University Press.} 
  \label{Sco1}
\end{figure}

\subsection{Group A}

Group~A  evidently coincides with the intermediate-age open cluster VdB-Hagen 202 \citep{vdb}. These authors describe it as a poor red cluster embedded in some nebulosity.  
It is by far the richest group detected in the covered area.  It has also the largest tangential velocity and the oldest age.
Our isochrone fit in fact yields an age of $500\,\mathrm{Myr}$, while both the astrometric and the photometric analysis support a heliocentric  distance of 1.65~kpc. This group has not been detected by
\citet{Damiani}, possibly because it does not contain young pre-main sequence M stars. On the other hand, close to this position \citet{Kuhn} found two rich groups (3 and 5, according to their numbering) 
of young stars slightly to the north of our group A, possibly our group F (see below).
In all cases, the large age (compared with the rest of the groups) and significantly diverse tangential motion lends support to a picture in which this cluster does not pertain to the association, but it is probably caught in the act of passing through it.  The groups D and E (see below) share the same properties of this group A.

\subsection{Group B}

This group is located in the southern edge of the H\,\textsc{ii} region G345.45+1.50, and appears very scattered. DBSCAN returns three different density peaks, that we indicate as B1, B2, and B3, but the area roughly corresponds to the very young star cluster Trumpler~24.  \citet{sege} also identified this group, which he indicated as 
group Trumpler~24~III. It is separated by a gap from the other groups identified in this study (see below). 
Besides the location, these three groups share the same age and kinematics.
In the literature, the distance to Trumpler~24 ranges from 1.6 to 2.2~kpc \citep{sege,Heske}. Our study favours the shortest distance, both from photometry and from \textit{Gaia} DR2 parallaxes. This group has clear counterparts both in \citet{Kuhn} and \citet{Damiani}.

\subsection{Group C}

	This is a very poor group located in the south-west corner  of the field we covered. It is essentially composed by early type stars (early B spectral type, judging from the colour-colour diagram), and therefore it shares the same age as Trumpler 24 (group B). Kinematics also is closer to Trumpler 24 than to VdB-Hagen 202.
		
\subsection{Group D}
        
In spite of its vicinity to the previous group C, this group appears to be significantly different. It does not contain very young stars, and its kinematics is closer to the star cluster VdB-Hagen 202 (group A) than to Trumpler~24 (gropus B1, B2, and B3). It seems also to be positioned somewhat in the background with respect to Trumpler 24. We propose that this group is a part of the group~A.

\begin{figure}[!t]
  \centering
  \includegraphics[width=\columnwidth]{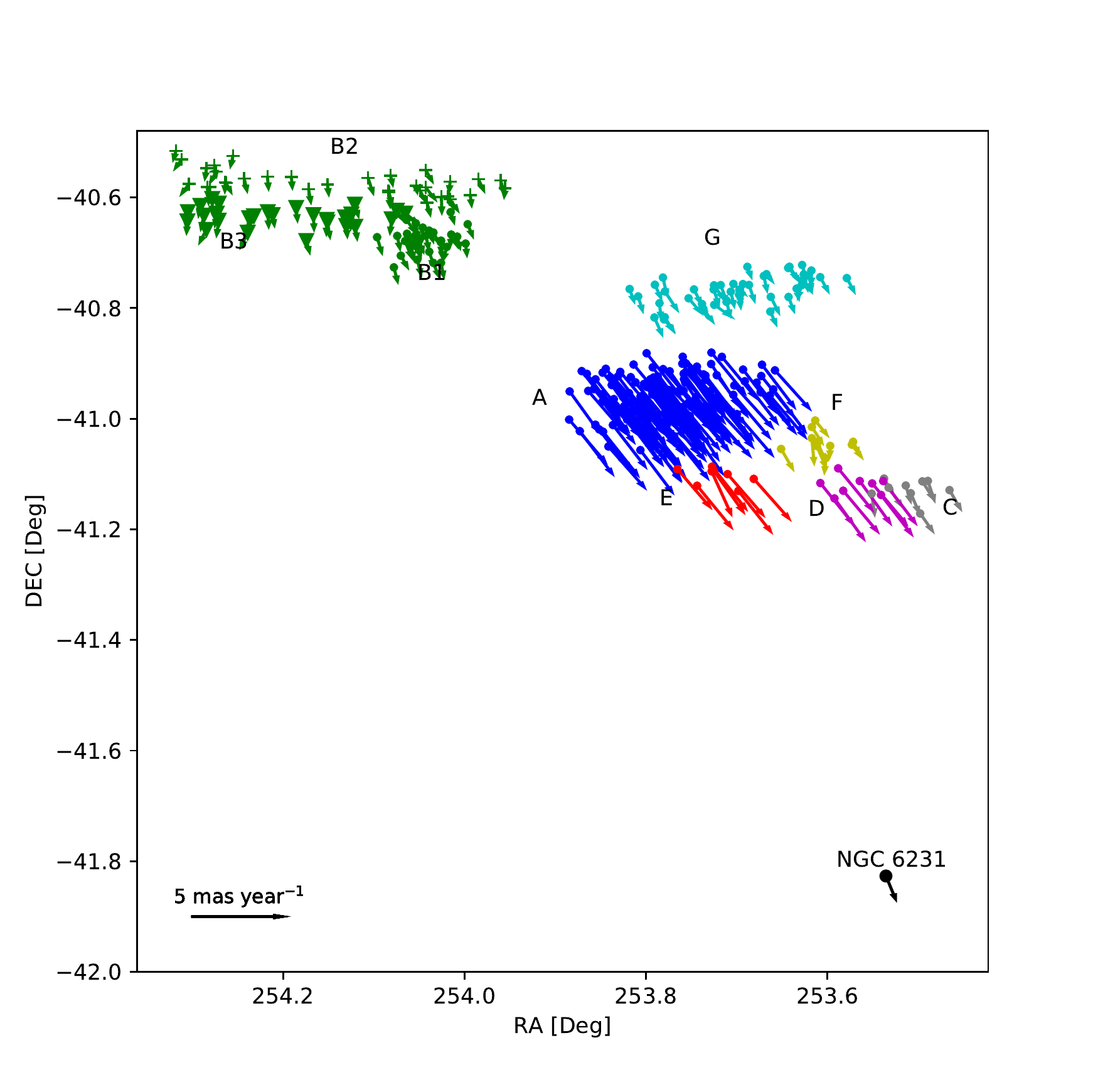}
  \caption{Tangential velocities for the various groups identified in the Sco~1 association.  In the lower right corner we indicate the tangential motion of NGC~6231  for comparison. Adapted from \cite{Yalialeva2020}, op.~cit.}
  \label{Sco2}
\end{figure}

\begin{figure}[!t]
  \centering
  \includegraphics[width=0.4\textwidth]{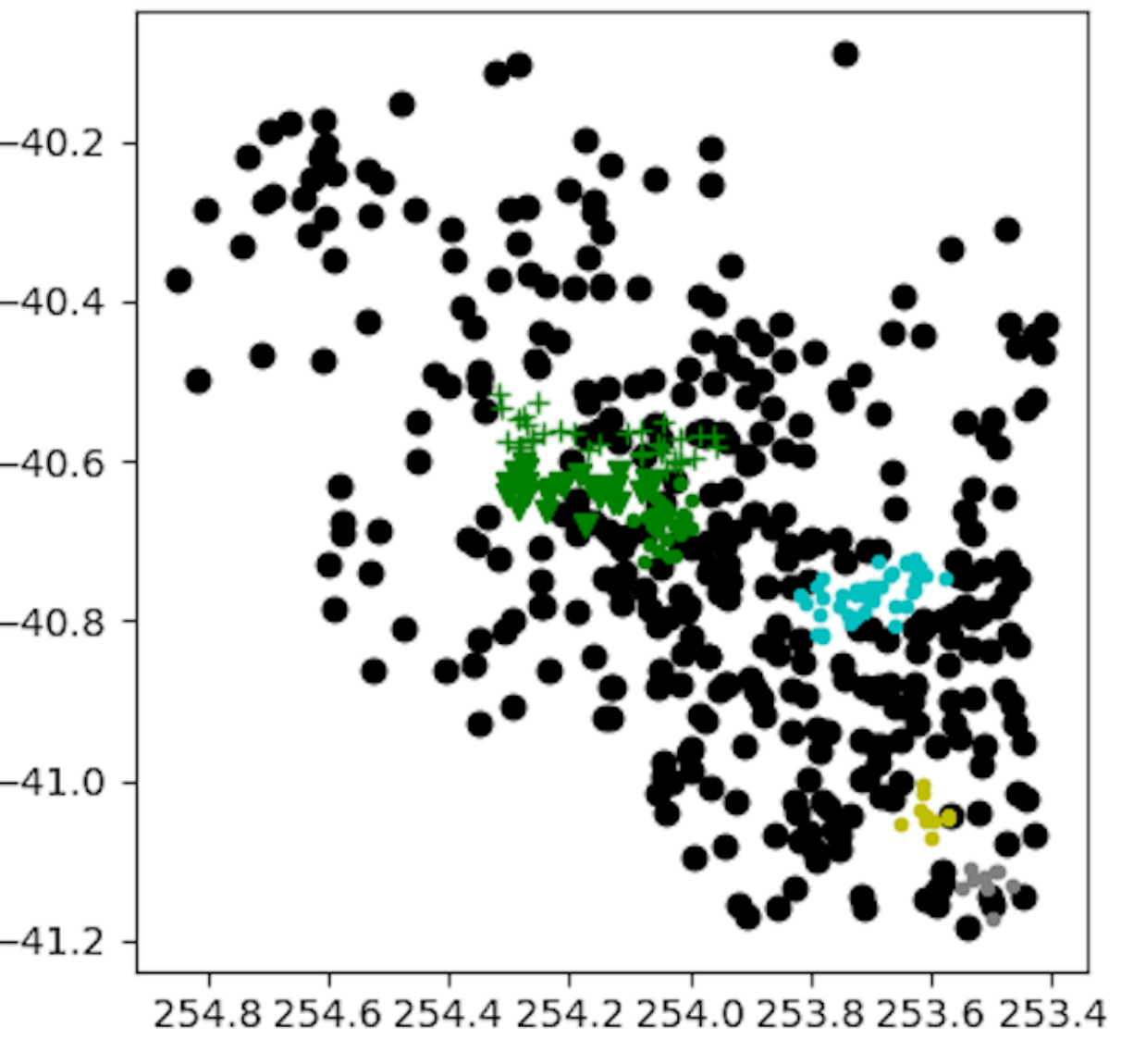}
    \includegraphics[width=0.4\textwidth]{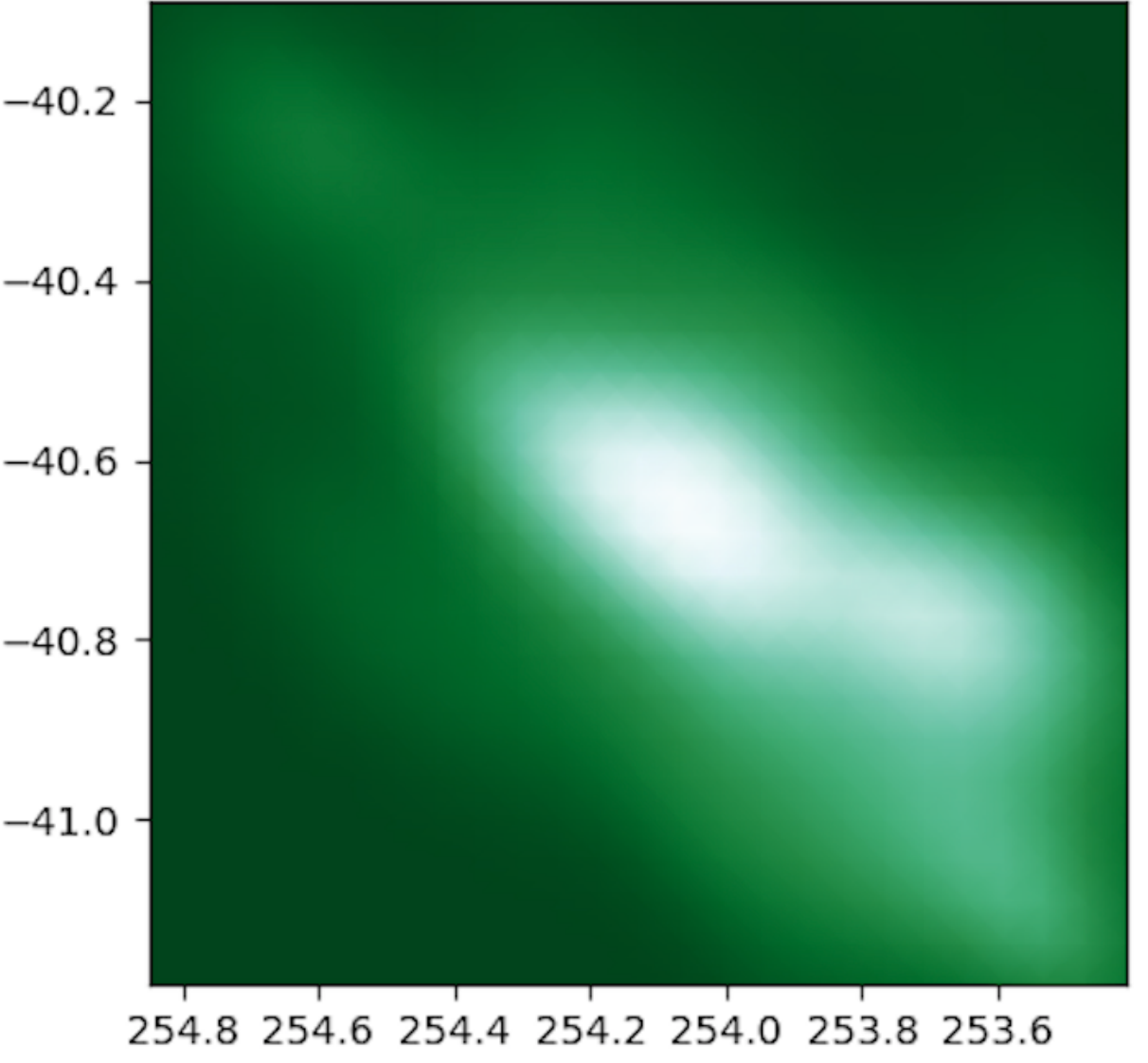}
      \caption{Distribution of family I stars (upper panel) and 2D density contour (lower panel). One can clearly notice two main concentrations separated by a density gap. Adapted from \cite{Yalialeva2020}, op.~cit.}
  \label{Sco3}
\end{figure}

\subsection{Group E}

This scarcely  populated group shares the same kinematics as the previous group D, and is situated very close to VdB-Hagen~202. Its paucity of stars prevents us from computing an accurate age
for the group. Nonetheless, it seems plausible to adopt an age close to that of group D. We will propose (see below) that this group, and group D as well,  are pieces of the group A (the stellar cluster VdB-Hagen~202).
	
\subsection{Group F}

This group possesses the same properties as group C.  It coincides with \citet{Kuhn} group 3. The few identified early type stars show age, kinematics, and distance consistent with Trumpler~24. It appears to coincide with \citet{sege} Trumpler~24~II group.
	
\subsection{Group G}

	This corresponds to \cite{sege} group Trumpler 24 II. Similar to the C and F groups, this group is young. Even if it is situated right to the north of the old A group (VdB-Hagen~202), it possesses mean proper motion components very close to Trumpler~24.

\subsection{Group families}
        
In full generality, the detailed analysis of the detected groups leads us to separate them in two different families:

	\begin{description}
	
	\item Family I: B (B1, B2 and B3), C, F and G groups have
colour-colour diagrams typical of a very young population,  and show the
presence of pre-main sequence stars in the colour-magnitude diagram. They share similar proper motion components: $\mu_{\alpha}= - 0.3\,\mathrm{mas\,yr}^{-1}$, $\mu_{\delta}= -1.3\,\mathrm{mas\,yr}^{-1}$.
	
	\item Family II: A, D and (maybe) E groups are significantly older and  have compatible proper motion components: $\mu_{\alpha}= -
1.7\,\mathrm{mas\,yr}^{-1}$, $\mu_{\delta}= - 3.7\,\mathrm{mas\,year}^{-1}$. They have distances on average larger than family I. 
	
	\end{description}

\noindent
Family II clearly has nothing to do with the Sco~1 OB association. It is quite reasonable to assume that the three groups are in fact just different portions of the very same star cluster, VdB-Hagen 202, which is undergoing tidal disruption while it is 
crossing the OB association. Fig.~\ref{Sco3} shows the distribution of family I stars. In the lower panel, two density peaks  are clearly visible, separated by a decrease in density, a sort of gap. It is tempting to associate this density gap to the recent passage of the old star cluster VdB-Hagen 202 through the Sco 1 association.

\noindent
In conclusion, also Sco1 OB association exhibits quite a complex structure, and it is not easy to depict a formation scenario for it. Here as well, as in the case of the previous examples, the lack of radial velocity data is limiting the study of the region significantly.

\section{Conclusions}
In this review, I discussed in some detail the two major theories of the formation of OB associations in the Milky Way: the monolithic collapse theory,  and the fractal theory of SF. These two theories predict different formation scenarios for the appearance of stellar (OB or T) associations.
 In the first case, an association is simply a dynamical stage of young star cluster evolution, caused by gas expulsion driven by massive stars evolution. In the second case, associations are the direct product of SF and therefore they reveal the ISM density peak distribution
 prior to the onset of the SF process.
 I then described in some detail observational material on three recently studied  OB associations, namely Cygnus~OB2, Vela~OB2, and Scorpius~OB1. Multi-colour
photometry are typically combined with \textit{Gaia} DR2 data (parallaxes and proper motion components) which are allowing the investigation of stellar OB
associations with much more detail than in the pre-\textit{Gaia} era. In all three cases the observational data outline the high complexity of these associations.
Several stellar groups are found with different spatial, kinematical and age properties. Two groups are found in Cygnus~OB2, six groups in Vela~OB2, and nine groups
in Scorpius~OB1. Interestingly, in these cases ages, spatial positions, and kinematics combine in very diverse fashions. 
 Based on these preliminary results it is tempting to conclude that SF occurred in a highly structured ISM, and involved molecular clouds with widely different masses and  differently distributed  (in number and mass) pre-stellar cores.
 These are of course preliminary conclusions, since many more OB associations await such detailed studies in the future. In any case, these studies show the
impressive impact that \textit{Gaia} is providing in the investigation of SF and OB associations in the Milky Way. It is anyway recommendable  to couple at some point
\textit{Gaia} data with precise radial velocity measurements of a sizeable number of stars, as the \textit{Gaia}-ESO \citep{GaiaESO} survey has been partially doing.
Only in this way, in fact, a real 3D kinematical study can be performed, and much more solid conclusions can be drawn on the origin and early evolution of stellar
associations. 

\begin{acknowledgement}
I sincerely thank the organisers of the XVI LARIM  conference for inviting me in Antofagasta to deliver this review. I also express my gratitude to Lidia Yalialeva
(Lomonosov Moscow University, Russian Federation)  and Rub\'en A. V\'azquez (Universidad de La Plata, Argentina) for their help in the assembly of this material. 
\end{acknowledgement}

\bibliographystyle{baaa}
\small
\bibliography{bibliografia}

\end{document}